\documentclass[conference]{IEEEtran}

\usepackage{graphicx}
\usepackage{xcolor}
\usepackage{amsmath}
\usepackage{enumitem}
\usepackage[printonlyused,withpage]{acronym}
\usepackage{algpseudocode,algorithm,algorithmicx}

\usepackage{comment}
\usepackage{hyperref}

\newcommand{\reg}{\textsuperscript{\textregistered}}
\newcommand{\intel}{Intel\reg}
\newcommand{\xeon}{Xeon\reg}

\usepackage{minted}
\setminted{fontsize=\footnotesize,baselinestretch=.97,linenos,frame=lines,xleftmargin=6pt,numbersep=3pt,mathescape=true,escapeinside=||,bgcolor=bg}
\definecolor{bg}{rgb}{0.97,0.97,0.97}
\newminted{cpp}{}

\newmintinline[cplusinl]{cpp}{}

\begin{document}

\pagestyle{plain}

\title{Julia GraphBLAS with Nonblocking Execution}

\author{\IEEEauthorblockN{
   Pascal Costanza\IEEEauthorrefmark{1},
   Timothy G. Mattson\IEEEauthorrefmark{2},
   Raye Kimmerer\IEEEauthorrefmark{3},
   Benjamin Brock\IEEEauthorrefmark{4}
}

\IEEEauthorblockA{\IEEEauthorrefmark{1}
  Independent researcher,
  Sint Truiden, Belgium}
\IEEEauthorblockA{\IEEEauthorrefmark{2}
  University of Bristol, Ocean Park, WA }
\IEEEauthorblockA{\IEEEauthorrefmark{3}
  National Energy Research Scientific Computing Center, Lawrence Berkeley National Laboratory, Berkeley, CA }
\IEEEauthorblockA{\IEEEauthorrefmark{4}
  Intel, Parallel Computing Lab,
  San Francisco, CA }
    }

\maketitle
\thispagestyle{plain}

\begin{abstract}
From the beginning, the GraphBLAS were designed for ``nonblocking execution''; i.e., calls to GraphBLAS methods return as soon as the arguments to the methods are validated and define a directed acyclic graph (DAG) of GraphBLAS operations.  This lets GraphBLAS implementations fuse functions, elide unneeded objects, exploit  parallelism, plus any additional DAG-preserving transformations.  GraphBLAS implementations exist that utilize nonblocking execution but with limited scope.  In this paper, we describe our work to implement  GraphBLAS with  support for aggressive nonblocking execution.  We show how features of the Julia programming language greatly simplify implementation of nonblocking execution.  This is \emph{work-in-progress} sufficient to show the potential for nonblocking execution and is limited to GraphBLAS methods required to support PageRank.
 \end{abstract}

\section{Introduction}

The initial definition of the C GraphBLAS API~\cite{bulucc2017design} specified a nonblocking execution model.  The function calls from a GraphBLAS library in program order define a Directed Acyclic Graph (DAG).  If a GraphBLAS library is initialized for nonblocking execution, the GraphBLAS implementation can utilize lazy evaluation, fusion of operations, or any other execution strategy that satisfies the semantics defined by the DAG.   In the GraphBLAS 2.1 specification~\cite{cSpec}, features that limited multithreaded execution were addressed so DAGs can be defined for multithreaded execution.

While limited forms of nonblocking execution have been implemented~\cite{SuiteSparseGraphBLAS,yzelmanNonblocking2022}, we are unaware of any implementation that utilizes compiler-based nonblocking execution.

In this paper, we report on our work to implement nonblocking GraphBLAS in Julia.  We call our implementation the applicative GraphBLAS or \emph{AppGrB}.  We use \emph{multi-stage programming}~\cite{Taha2004}  to generate a symbolic representation of the code at runtime, dynamically compile it, and then execute it. Julia’s support for multi-stage programming was essential for this work. It lets us generate and compile code from a first-class representation of a GraphBLAS method tree at runtime and support aggressive inlining and fusion into tight loops.

\section{Nonblocking execution}

The GraphBLAS C API~\cite{buluc2019graphblas} defines a series of \textit{GraphBLAS operations} that act on matrix, vector and scalar objects.  The ordered sequence of GraphBLAS operations in the program (in \emph{program order}) define a directed acyclic graph (DAG) with nodes as GraphBLAS operations and edges as dependencies between operations.  When initializing a GraphBLAS library, a user may select one of two execution modes:

\begin{itemize}
  \item {\bf Blocking mode}: When any GraphBLAS operation returns, its execution is complete, any side effects are fully resolved, and associated GraphBLAS objects are fully materialized.
  \item {\bf Nonblocking mode}: GraphBLAS operations may return once the input arguments have been validated but before computations have begun.  Objects communicated through edges, though computation is pending, are still available for use in subsequent operations in the DAG.
\end{itemize}
\noindent
Nonblocking execution provides flexibility needed to optimize the execution of the DAG.  Execution may be deferred  until the full DAG has been defined allowing rewrite rules to fuse operations, extract parallelism, elide unused objects or intermediates, and other approaches to optimize the DAG.   The result of the DAG's execution must be the same in blocking and nonblocking modes (other than effects such as nonassociativity due to rounding in IEEE-754 arithmetic).

\section{Nonblocking Execution in Julia}

AppGrB uses \emph{multi-stage programming}~\cite{Taha2004} to generate a symbolic representation of source code and dynamically compile it at runtime prior to execution.   A good example of multi-stage programming is the streaming computations in the Strymonas library \cite{Kiselyov2017}. Some programming languages directly support multi-stage programming, including BER MetaOCaml, Scala, Common Lisp, and Julia.  In Julia, code can be generated at runtime through \emph{quoting} and \emph{interpolation}, with the \texttt{eval} function compiling the code at runtime. In AppGrB, we maintain an explicit representation of the DAG of GraphBLAS methods.  When compilation is necessary, the DAG is transformed into a symbolic code representation, JIT compiled, and executed.  This happens when the GraphBLAS \texttt{wait} method is called, which ensures materialization of a GraphBLAS object. %This is when AppGrB performs code generation, dynamic compilation, and execution of the dynamically compiled code.

%\bab{The language above previously said ``as well as methods that produce an opaque result, which implicitly require materialization.''  I believe you mean produce a non-opaque result, like getElement, but please double-check me and correct if I'm wrong here.}
%\pascal{I removed the language around ``opaque'' methods, because it matters much less than I originally thought when I started AppGrB. getElement won't trigger materialization, but just computes the respective element, nothing else. Even reductions to scalars produce a Scalar object that still needs to be materialized to access its value. So it's easier to just drop this tiny little detail.}

%\pascal{This is probably my fault, but there seems to be a misunderstanding that became apparent in the rewritten text (now changed): Every GraphBLAS method invocation ``just'' instantiates a struct that represents that method, with references to other such structs representing the inputs, which may themselves be instances of structs that represent other methods, or materialized representantions as leaves. The ``symbolic'' representation is the code generated from these method trees -- not as text, but as data structures, hence ``symbolic'' -- that can then be compiled into machine code. This symbolic representation is a standard feature of Julia, and can be generated with quotation and interpolation. Only the code generation and compilation is delayed, the DAGs are immediately returned from each GraphBLAS method call.}

\begin{figure*}[ht!]
\hrule
	\caption{The generated inner loop for \texttt{z = ewise\_mult(*, x, y)}.}
\label{Fig:loop1}
\footnotesize
\begin{verbatim}
  for col_1 = 1:nof_cols_2
      result_3[col_1] = vector_ref_4.values[col_1] * vector_ref_5.values[col_1]
  end
\end{verbatim}
\hrule
\end{figure*}

\begin{figure*}[ht!]
\hrule
	\caption{The generated inner loop for \texttt{z = ewise\_mult(*, x, apply((x) -\> x + 1), y)}.}
\label{Fig:loop2}
\footnotesize
\begin{verbatim}
  for col_1 = 1:nof_cols_2
      result_3[col_1] = vector_ref_4.values[col_1] * ((x) -> x + 1)(vector_ref_5.values[col_1])
  end
\end{verbatim}
\hrule
\end{figure*}

Consider GraphBLAS code to multiply two vectors \texttt{z = ewise\_mult(*, x, y)}. The inner loop of the generated code is shown in Figure~\ref{Fig:loop1}.\footnote{Variable names are generated symbols with numbers attached to them to prevent accidental name captures. Our presentation is simplified  by avoiding this and other low-level details.}.  As another example, consider a  modified piece of GraphBLAS code that  adds one to each entry of the second input vector \texttt{z = ewise\_mult(*, x, apply((x) -> x+1, y))}. The inner loop of the generated code is shown in Figure~\ref{Fig:loop2}.
In this case, the inline lambda expression is immediately applied to an argument. The compiler can easily optimize away that function call.  This technique is well-known for streaming abstractions~\cite{Kiselyov2017} and can be applied across collections of GraphBLAS methods.  We act on two additional observations about efficient JIT-based nonblocking execution.
\begin{enumerate}
  \item For best performance, we must minimize dynamic compilation, especially in inner loops. Therefore, JIT-compiled kernels must be cached and reused. We define \emph{GraphBLAS method signatures} as keys into the cache to find these previously compiled kernels.
  %for such caches that can be extracted from a GraphBLAS method DAG.
  \item  We need efficient algorithms for a particular kernel. For example, in element-wise multiplication it is best to  loop over the indices of the matrix with the fewest non-zero elements. However, the sparsity of the matrices is not known when generating code.   Worse, if only one of the matrices is materialized, it may not be wise to iterate over its indices, since the other matrix might end up having fewer non-zero elements.
Such issues also arise for masked computations since the masks themselves may not yet be materialized. To drive algorithm selection, we therefore use \textit{estimated fill ratios} based on quick computations on the representation of a GraphBLAS method.
%GraphBLAS method (sub-)DAG.
\end{enumerate}
We provide more details about the general approach of AppGrB, multi-stage programming, caching of compiled method representations, and estimated fill ratios below.

\subsection{General approach}

In AppGrB, users store data in \texttt{GrBScalar\{T\}}, \texttt{GrBVector\{T, I\}}, and \texttt{GrBMatrix\{T, I\}} containers, where \texttt{T} is the scalar type of values stored in the container and \texttt{I} is the integer type used for indices.  These user-facing containers may contain either the materialized matrix itself or a tree representation of the method tree that will later be used to produce the matrix.
This is implemented as a struct that contains a reference to a subtype of \texttt{AbstractGrBScalar\{T\}},  \texttt{AbstractGrBVector\{T, I\}}, or \texttt{AbstractGrBMatrix\{T, I\}}, respectively.
For example for vectors, materialized subtypes include \texttt{SparseVector\{T, I\}}, \texttt{BitSetVector\{T, I\}} and \texttt{FullVector\{T, I\}}.

GraphBLAS methods are not executed immediately, but are delayed to support nonblocking execution. To facilitate this, GraphBLAS methods in AppGrB do not have side effects, unlike in the GraphBLAS C specification where results are destructively assigned to one of the method parameters. Instead, GraphBLAS methods in AppGrB return new subtypes of the \texttt{AbstractGrB} container types that represent the computation involved in the method.
%, all wrapped within the appropriate user-visible container representation.

\begin{figure}[ht!]
\hrule
	\caption{The \texttt{MxV} Julia struct.}
\label{Fig:mxv}
\footnotesize
\begin{verbatim}
struct MxV{DC, DA, DB, Index <: Integer}
                     <: AbstractGrBVector{DC, Index}
    ncols::Int
    A::GrBMatrix{DA, Index}
    B::GrBVector{DB, Index}
    identity::DC
    add::Any
    mul::Any
end
\end{verbatim}
\hrule
\end{figure}

For example, when calling \texttt{mxv} with an identity value, add and multiply operators, and two input containers, an instance of the struct \texttt{MxV\{T, I\}} is returned, which is a subtype of \texttt{AbstractGrBVector\{T, I\}}, as shown in Figure~\ref{Fig:mxv}. This instance has fields that refer to the identity value, the operators and the two input containers, alongside other relevant information, wrapped in a \texttt{GrBVector\{T, I\}}.

GraphBLAS methods can be called both on materialized and non-materialized containers, forming trees with a particular (non-materialized) method representation as its root, and materialized container representations as their leaves. When materialization of such a method tree is requested by calling \texttt{wait}, that tree is translated to a symbolic representation of source code, dynamically compiled, and then executed, as described below.

\subsection{Multi-stage programming}

%\bab{In my opinion, it would be best to cut most of this multi-stage programming section and instead focus a little more on how the DAGs are actually built in AppGrB (expanding the last paragraph).  I'm happy to help write this, but I'm not sure I'd get all the technical details write.  In particular, it would be nice to see 1) what are the different expressions that you have types for (MxV, Ewise, ?), 2) precisely the conditions under which these expressions will have to be materialized, and 3) how code is generated from an expression.  Can we write, at least in prose at a high level, the algorithm that iterates over the DAG and generates the code?  An example would be nice.}

Multi-stage programming is a technique for metaprogramming, where code is generated at runtime, dynamically compiled, and then executed. To support multi-stage programming, there must be a way to symbolically represent code, and mechanisms to construct such representations. Python's \texttt{eval} could be regarded as a very simple form of multi-stage programming. The code is represented as a string, and Python's \texttt{eval} function takes care of translating the code to bytecode and executing that bytecode. However, that approach is not very sophisticated: The code representation is not symbolic, which makes code construction cumbersome.

Proper multi-stage programming uses some form of \textit{quotation} (sometimes called \text{quasiquotation} \cite{Bawden99}) to construct and represent code, which can be understood as a form of code templates. For example, in Julia, \texttt{:(2 + 2)} is quoted code that represents the computation of adding two numbers. \textit{Interpolation} (sometimes  called \textit{splicing} \cite{Bawden99})
lets us insert values into such quoted code. For example, \texttt{:(2 + \$x)} represents code that adds a number to whatever value \texttt{x} is bound to at the time this code is constructed. Each piece of quoted code is a first-class value that can be printed, bound to variables, inspected and passed to functions for further code construction. In order to execute the represented code, Julia's \texttt{eval} function compiles the translated code into machine language (using LLVM) and then executes it.  To compile a piece of code once, but then execute it multiple times, the quoted code can be represented as a lambda expression. Consider the following fragment:

\begin{verbatim}
x = 42
f = eval(:(() -> print($x)))
f()  # prints 42
x = 11
f()  # prints 42
\end{verbatim}

Note that, since the value of \texttt{x} was interpolated, the constructed code prints the literal value \texttt{42}, no matter which value is subsequently assigned to \texttt{x}. To ensure that the code always prints the current value of \texttt{x}, it should not be interpolated (i.e., it should mention \texttt{x} without the preceding \texttt{\$}). In other words, careful nesting of quotation and interpolation precisely annotates  which parts of the code should be executed at which \textit{stage}: during code construction, or at later code execution.

AppGrB  method trees are converted into quoted code using multi-stage programming, where each subtype of the various \texttt{AbstractGrB} types contribute their own logic to the resulting code, including both materialized and non-materialized representations. An invocation of \texttt{wait} thus typically generates one or two outer loops to produce the result container, and the nested methods are fused into that loop.

\begin{figure*}[ht!]
\hrule
	\caption{Implementations of \texttt{foreach\_entry} for full and sparse materialized vectors, and for the GraphBLAS apply method.}
\label{Fig:foreach_entry_full_sparse}
\footnotesize
\begin{verbatim}
function foreach_entry(vec::FullVector{T, Index}, vec_name, block) where {T, Index <: Integer}
  @gensym vec_ref col
  :(let $vec_ref::FullVector{$T, $Index} = $vec_name.ref
      for $col in 1:$vec_ref.ncols
        $(block(col, :($vec_ref.values[$col])))
      end
    end)
end

function foreach_entry(vec::SparseVector{T, Index}, vec_name, block) where {T, Index <: Integer}
   @gensym vec_ref index
   :(let $vec_ref::SparseVector{$T, $Index} = $vec_name.ref
       for $index in eachindex($vec_ref.indices)
         $(block(:($vec_ref.indices[$index]), :($vec_ref.values[$index])))
       end
     end)
end

function foreach_entry(vec::VectorApply{Out, In, Index}, vec_name, block) where {Out, In, Index <: Integer}
   @gensym vec_ref vec_apply_base
   :(let $vec_ref::VectorApply{$Out, $In, $Index} = $vec_name.ref,
         $vec_apply_base = $vec_ref.base
       $(foreach_entry(vec.base, vec_apply_base, (col, val) -> block(col, :($(vec.op)($val)))))
     end)
end
\end{verbatim}
\hrule
\end{figure*}

\begin{figure*}[ht]
\hrule
	\caption{Materialization of a GraphBLAS vector.}
\label{Fig:materialization}
\footnotesize
\begin{verbatim}
function materialize(vector::AbstractGrBVector{T, Index}, ref) where {T, Index <: Integer}
  compiled = compile(signature(vector), function ()
                       @gensym vec_name indices values
                       :(($vec_name) ->
                            let $indices = Vector{$Index}(),
                                $values = Vector{$T}()
                              $(foreach_entry(vector, vec_name, (col, val) -> :(begin
                                                                                  push!($indices, $col)
                                                                                  push!($values, $val)
                                                                                end)))
                              SparseVector{$T, $Index}($indices, $values)
                            end)
                     end)
  @invokelatest compiled(ref)
end
\end{verbatim}
\hrule
\end{figure*}

Each subtype of the  \texttt{AbstractGrB} types must implement a set of functions for expressing  ways to iterate over the elements of the container. For example, to iterate over the non-zero entries, they must implement a \texttt{foreach\_entry} function. However, instead of performing the iteration directly, it generates a symbolic representation of the iteration. See Figure~\ref{Fig:foreach_entry_full_sparse} for implementations of \texttt{foreach\_entry} for both full and sparse vectors, as well as the GraphBLAS \texttt{apply} method. The \texttt{foreach\_entry} function receives the container for which to specialize the generated code, the name of the container in the surrounding code in which the generated code needs to be embedded, and a \texttt{block} function that gets passed expressions for the respective column and value in each iteration. The latter \texttt{block} function can then generate further code. Note how the \texttt{foreach\_entry} method for \texttt{VectorApply} recursively invokes \texttt{foreach\_entry} on its \textit{base} --- i.e., on the vector to whose values the respective operator should be applied --- and how the function passed to that recursive invocation wraps the respective value from the inner iterator and then invokes the \texttt{block} function that the outer iterator received.

In Figure~\ref{Fig:materialization}, we show an example of how this iterator protocol is used to materialize an arbitrary GraphBLAS method tree that results in a vector. It calls an (AppGrB-defined) \texttt{compile} function with the signature of the vector, and a function that generates the symbolic code representation that can eventually return a materialized vector. That code is a lambda expression that expects the vector method tree, sets up the low-level index and value vectors, has a loop fused into its code as generated by \texttt{foreach\_entry}, and finally creates a sparse GraphBLAS vector.

Note that the code presented here has been simplified. There are many more options for iteration, including  iterating only over the indices but not the values, iterating only over the values, iterating step-by-step (with \texttt{first} and \texttt{next} methods) to alternate between two input containers (such as for element-wise addition), and random accesses for certain special cases of multiplication. Also what is presented here as a single \texttt{foreach\_entry} function is split up into setup, iteration, and more fine-grained index and value accessors, including accessing an iso value only once for iso-valued containers. AppGrB also allows for both sequential and parallel iteration, which requires generating different forms of loops.

\subsection{Caching of compiled method trees}

Multi-stage programming can be used to generate code at runtime for a particular GraphBLAS method tree.  The generated code takes runtime properties of the method call into account such as the storage format of arguments passed to the method (e.g., whether they are sparse or dense),  whether they are iso-valued (i.e., all defined elements of the object have the same value), the concrete operations associated with the monoids or semirings (which can be inlined as part of code construction), and so on.

%%% Commented out signature details to make the paper fit for HPEC
%\begin{figure*}[ht!]
%\hrule
%	\caption{The signature method for \texttt{MxV}.}
%\label{Fig:mxvsig}
%\footnotesize
%\begin{verbatim}
%signature(vector::MxV{DC, DA, DB, Index}) where {DC, DA, DB, Index <: Integer} =
%    (:mxv, signature(vector.A), signature(vector.B), vector.identity, vector.add, vector.mul,
%     iso(vector.A), iso(vector.B), isrowdense(vector.A), iscoldense(vector.B), DC)
%\end{verbatim}
%\hrule
%\end{figure*}

However, when a program calls \texttt{wait} on a method tree, such as the one for \texttt{mxv} in Figure~\ref{Fig:mxv}, AppGrB will first query the cache and retrieve a previously compiled kernel if it exists. To facilitate this, we generate a recursive generic \texttt{signature} function that computes the key for such a lookup. %Its definition for a matrix-vector multiplication is shown in Figure~\ref{Fig:mxvsig}.
%This signature encodes the information as a tuple which the code construction takes into account as it generates optimized code, including whether the involved containers are iso-valued or not, whether each row of the matrix \texttt{A} has at least one value, and whether the vector \texttt{B} is dense.
The constructed code is a lambda expression that expects an instance of the associated struct (as shown in Figure~\ref{Fig:mxv}), and will only work for instances that have the same signature.

\subsection{Estimated fill ratios}

In general, the sparsity (number of non-zeros) of the result of a GraphBLAS method cannot be known in advance.
When we generate code for, say, multiplying two containers element-wise, it is more efficient to let the sparser input container guide the overall iteration. However, in AppGrB, the input containers themselves may be inner nodes of a GraphBLAS method tree, so the sparsity of input containers may not be readily available. The only way to reliably infer the sparsity of the result of a GraphBLAS method is to execute it immediately, which defeats the purpose of nonblocking execution. However, approaches exist to predict the sparsity of the result of a matrix operation with different degrees of accuracy~\cite{Cohen1997, Amossen2014, Sommer2019, Du2022}. These predictors are traditionally used to estimate output array sizes, but we use them here to drive algorithm selection. As a first step, we opted for \emph{naive metadata estimators}~\cite{Sommer2019}, which are based on apparent properties of input data. For example, when multiplying two input containers, the estimated sparsity of the result is just the product of the sparsities of the input containers (in percentage of the container size). Other GraphBLAS methods lead to similarly straightforward estimators. For materialized containers, the “estimated” sparsity is just the number of non zeros divided by the container size. There are  exceptions. Some GraphBLAS select methods use an arbitrary user function to determine which elements of an input container are retained in the result. For predefined selection functions, it is also possible to provide a ``naive'' estimator. In general, we foresee that users may need the option to define their own estimators.

Currently, we use such estimators as follows:
\begin{itemize}
\item When generating output vectors, we use the estimator to determine whether the representation of the result is sparse, bitset, or full.
\item In element-wise multiplications, the estimator determines which of the two input containers drives the multiplication loop.
\item In element-wise additions, the estimator determines whether to use an outer loop that ranges over the full dimensions (when the result is expected to be dense or almost dense), or whether to iterate over the indices of the input containers.
\end{itemize}
We expect to use estimators for masked operations, and we already use them to preallocate output arrays.
It is important to stress that the estimators have no influence on the correctness of the involved algorithms. When an estimator is incorrect, it will at worst  have a negative impact on performance.
%In future work we will evaluate other estimators in the literature to see if more costly estimators would significantly improve performance.

%these are preliminary results superceded by subsequent results.
% there are some key points of discussion in this file so we should keep it around ... just not in the paper.
%\input{benchmarks}

\section{Performance Results}

To characterize the performance of our nonblocking implementation of the GraphBLAS in Julia, we considered the PageRank algorithm  with several GraphBLAS systems.
\begin{itemize}
\item {\bf AppGrB nonblocking}: The nonblocking GraphBLAS described in this paper using Julia.
\item {\bf AppGrB blocking}: The AppGrB nonblocking code forced to execute in blocking mode by calling \texttt{GrB\_wait()} after each operation; i.e., force the system to wait for each operation to finish and for every GraphBLAS object to be complete, thus matching the definition of the blocking mode in GraphBLAS.
\item {\bf AppGrb C++ stub}: To understand the quality of the LLVM code generated by Julia, we took the generated Julia code and hand-implemented it using C++.
\item {\bf SuiteSparse}:  We used the SuiteSparse implementation of the GraphBLAS version 10.1.0 and the PageRank code from LAGraph version 1.1.4 (which is based on the algorithm in~\cite{szarnyas2021}).
\end{itemize}

\begin{figure}[ht!]
\hrule
\caption{The PageRank implementation in AppGrB.}
\label{Fig:appgrb-pagerank}
\footnotesize
\begin{verbatim}
function pagerank_gap(AT::GrBMatrix{Float32, Index},
                      out_degree,
                      damping,
                      tolerance,
                      itermax)
                            where {Index <: Integer}
    n = nrows(AT)
    scaled_damping = (1 - damping) / n
    teleport = scaled_damping
    rdiff = GrBScalar(1.0f0)
    t = GrBVector{Float32, Index}(n)
    r = GrBVector{Float32, Index}(n, 1.0f0 / n)
    d = GrBVector{Float32, Index}(n, 1 / damping)
    d = ewise_add(:((x, y) -> max(x / $damping, y)),
                  conv(Float32, out_degree), d)
    wait(d; parallel=true)
    rt = GrBVector{Float32, Index}(n, teleport)
    iter = 1
    while iter <= itermax && rdiff() > tolerance
        t = r
        r = ewise_mult(:(/), t, d)
        r = mxv(0.0f0, :(+), :((_, x) -> x), AT, r)
        r = ewise_add(:(+), rt, r)
        r = wait(r; parallel=true)
        t = ewise_add(:((x, y) -> abs(x - y)), t, r)
        rdiff = reduce(:(+), 0.0f0, t)
        iter += 1
    end
    (r, iter)
end
\end{verbatim}
\hrule
\end{figure}

Benchmarks were run on a system equipped with two \intel{} \xeon{} Platinum 8368 processors and 503 GB of memory.  Each processor has 16 performance cores and 22 efficiency cores for a total of 76 cores across both processors.  Julia programs go through the following  steps when calling wait on a GraphBLAS method tree.
\begin{enumerate}
\item Determine the method signature
\item Use the method signature to look up a possibly previously compiled code from the compilation cache.
\item If the previously compiled code does not exist, generate the symbolic representation, compile it to machine code, and add it to the compilation cache.
\item Execute the compiled code.
\end{enumerate}

%\bab{I filled in the processor details below.  It opens a bit of a can of worms, since it's one of the new split P-core/E-core systems, meaning some of the cores are faster than others.  That could definitely explain why Pascal observed some wonky behavior using different versions of OpenMP (I would expect a split P/E-core execution to be quite sensitive to changes in scheduling.)  Not sure if we want to get into that whole can of worms; probably not.}

%\tim{ oh dear, the P/E core issue does complicate things.  it explains why dynamic load balancing is so advantageous.  For a short paper, I think we need to just avoid the details raised by this point}.

We ran programs once to fill the code caches, a so-called \emph{warmup} run, and then report the average runtime from 16 additional runs.   During the warmup run, all 4 steps defined above execute, however in subsequent runs, step 3 is  skipped.  Ignoring this cost (about three seconds) gives us a more consistent way to compare appGrB execution to other approaches but is often justified since in practice, a GraphBLAS method is called many times in a single program making the one time cost of step 3 insignificant. In future work, we will investigate use of a persistent cache to hold compiled code between runs for use with recurrent signatures.

%\pascal{Yes, indices are 32 bit in the Julia and C++ runs. The data set is GAP-web.mtx from the GAP benchmark suite. The element type is 'integer'. The compiler is Intel(R) oneAPI DPC++/C++ Compiler 2025.0.1 (2025.0.1.20241113). I just called ``make'' both in the GraphBLAS and LAGraph repositories, no special settings. The processor is https://www.intel.com/content/www/us/en/products/sku/212455/intel-xeon-platinum-8368-processor-57m-cache-2-40-ghz/specifications.html (x 2).}

All four programs use 32 bit indices.  These are sufficient to support the test cases defined in the GAP benchmark~\cite{GapStudy2020}.  Although in the general case, the SuiteSparse GraphBLAS matches the GraphBLAS C API specification requirements with 64 bit indices, it automatically reduces indices to 32 bit when matrix and vector dimensions are within range.  This is a recent optimization added in version 10 of the SuiteSparse GraphBLAS.

\begin{figure*}[ht]
\centering
%\centerline{\includegraphics{PerfBarChart}}
\includegraphics[scale=0.13]{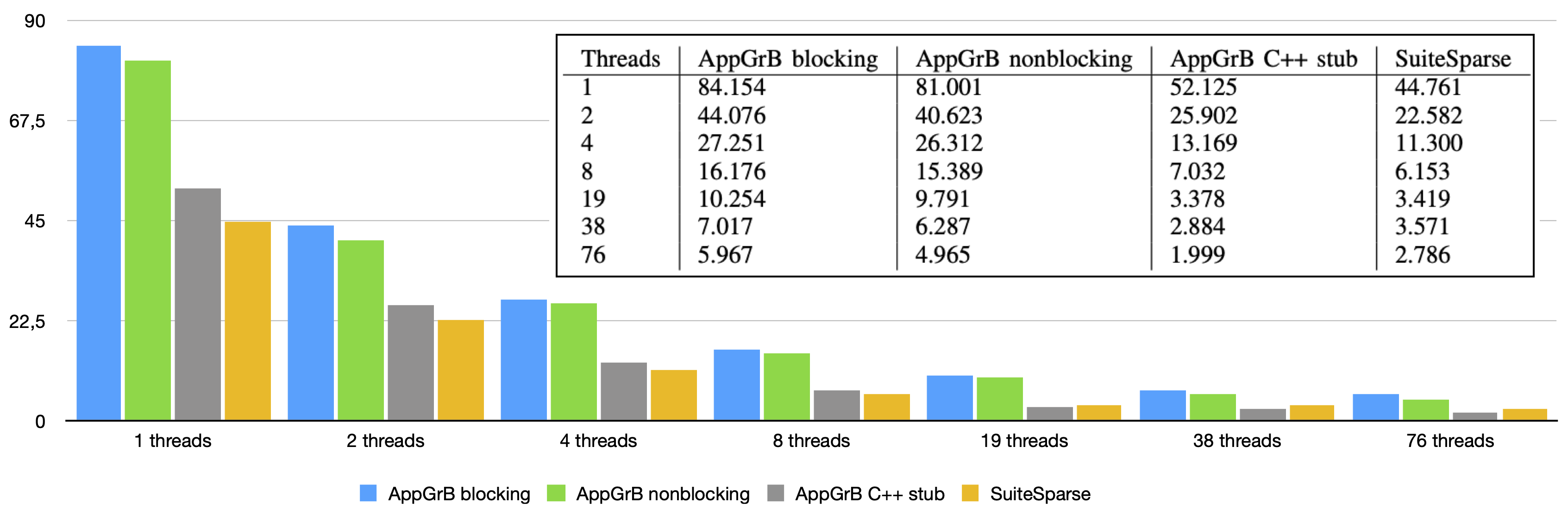}
\caption
{\textbf{Page Rank Performance} -- \small
Average of 16 runs in seconds for PageRank and the GAP-web.mtx data set.}
\label{figure:ChartTab}
\end{figure*}

The results are presented in Figure~\ref{figure:ChartTab}.  We used the GAP-web.mtx data set~\cite{Beamer:GAP} which is a directed graph with 50.6 million vertices and 1.9494 billion edges.  It has substantial locality and has a high average degree across the vertices.

%% \begin{table}[]
%%    \centering
%%    \resizebox{\linewidth}{!}{
%%    \begin{tabular}{l | l | l | l | l }
%%    Threads  & AppGrB blocking & AppGrB nonblocking & AppGrB C++ stub & SuiteSparse\\
%%        \hline
%%      1  & 84.154 & 81.001 & 52.125 & 44.761 \\
%%      2  & 44.076 & 40.623 & 25.902 & 22.582 \\
%%      4  & 27.251 & 26.312 & 13.169 & 11.300 \\
%%      8  & 16.176 & 15.389 & 7.032  & 6.153 \\
%%      19 & 10.254 & 9.791  & 3.378  & 3.419 \\
%%      38 & 7.017  & 6.287  & 2.884  & 3.571 \\
%%      76 & 5.967  & 4.965  & 1.999  & 2.786 \\
%%    \end{tabular}
%%  }
%%    \vspace{0.1in}
%%    \caption{Page Rank results for the GAP-web.mtx data set.  Times in seconds  are average of NN runs and do not include JIT compilation time}
%%    \vspace{-0.1in}
%%    \label{tab:perf}
%% \end{table}

Comparing the Julia results (\emph{AppGrb blocking} and \emph{AppGrb nonblocking}) to the  C++ stub code we see there is considerable room for improvement in the LLVM code generated by the Julia backend.
For high thread counts (76, 38 and 19), the best results are with the C++ stub code.  For lower thread counts, SuiteSparse GraphBLAS performs better.  Parallel efficiency is 34 percent for the AppGrB C++ stub at 76 threads, and 21 percent for SuiteSparse GraphBLAS.

AppGrB with nonblocking execution is noticeably faster than the same code running in blocking mode (AppGrB blocking).   This demonstrates the overall value of our approach for nonblocking execution in the GraphBLAS.  Not only was the performance at all thread counts consistenly better, the parallel efficiency at 76 threads was 0.21 for nonblocking mode compared to 0.18 for blocking mode.

%parallel efficiency ... runtime(1 thead)/runtime(76 threads)* (1.0/76)
%parallel efficiency suitesparce = 0.20
%parallel efficiency AppGrB C++ stub = 0.11

\section{Related Work}
The ALP/GraphBLAS library~\cite{yzelmanNonblocking2022} supports tiling and parallel execution of element-wise operations when executing in nonblocking mode.  ALP/GraphBLAS defers execution and produces a chain of objects representing operations and its iteration space.  Once objects are materialized, the runtime iterates over the iteration space in tiles, applying all operations within a tile before moving to the next.  This improves cache efficiency, but it does not assemble the whole DAG before generating fused kernels. Instead, it relies on lambda expressions to structure the tiled execution, with SpMV and SpMM forcing materialization.
ALP/GraphBLAS also dynamically analyzes dependencies between operations, placing them across queues that can be executed in parallel.  Operations with no dependencies can be placed in a new queue, while operations with dependencies are placed in a pre-existing queue, merging queues where necessary.

For sparse linear algebra, updates and deletions of values into sparse arrays can add a great deal of overhead.  In the SuiteSparse implementation of the GraphBLAS, as described in~\cite{davis2023algorithm,SuiteSparseGraphBLAS}, nonblocking mode is used to reduce this overhead. When a value is to be deleted, it is marked as a \emph{zombie}.  The value remains inside the sparse data structures, but removal is delayed.  Likewise when a value is added to an empty location in a sparse matrix, it can be marked as a \emph{pending tuple} making it available for use in GraphBLAS operations but without committing the update in the sparse array itself.  These modifications to the sparse data structures are deferred so they can occur later in a more optimal way.  On a larger scale, there are times when operations produce jumbled sets of tuples representing a sparse array.  Rather than immediately sorting these into one of the SuiteSparse storage formats, the sort can be deferred or in some cases even elided.  SuiteSparse GraphBLAS uses multi-stage programming in its JIT to optimize user-defined operators.  When user-defined operators are used, the JIT will compile new kernels generated by pasting a string representing the user-defined binary operator and/or monoid into a kernel skeleton, using an external file compiler.  This JITed kernel, generated with the user-defined operator inline, is then launched instead of using a pre-defined kernel that calls the operator through a function pointer, which is often slower.  Additional optimizations that utilize nonblocking execution are an ongoing effort in the SuiteSparse GraphBLAS project.
%from Tim Davis .... I forgot to mention one thing:  another pending work in nonblocking mode is "jumbled".  Many algorithms naturally produce CSR / CSC matrices with their indices in sorted order, in each row of a CSR matrix.  Some do not.  For those that do not, I allow them to produce their output with column indices possibly out of order.  I call these "jumbled" matrices.  This "jumbled" status is pending work that is only done when needed, or by GrB_wait.Many algorithms don't care if their inputs are jumbled, so I can sometimes skip this work entirely.  Algorithms which don't tolerate jumbled inputs must do the pending work and sort each row of a CSR matrix, or hypersparse CSR matrix.

GBTLX~\cite{raogGbtlx2020} uses SPIRAL to generate optimized C code from an execution trace gathered from a GBTL GraphBLAS program.  While GBTLX does fuse operations together into a single kernel in its generated trace, it does this only for blocking GraphBLAS programs, meaning a user cannot take advantage of GraphBLAS nonblocking mode with GBTLX.

PyGB~\cite{chamberlinPygb2018}, a Python interface for GraphBLAS, defers execution by building an expression tree, analogous to our method trees, that represents the computation necessary to obtain each GraphBLAS object.  When an expression tree must be materialized, it is used to generate a GBTL program, which is then executed.  While expression trees could in theory be fused and executed in a parallel fashion, GBTL currently has no support for nonblocking mode and executes the expression tree one operation at a time without fusion.

The Julia library GraphBLAS.jl~\cite{kimmerer2024} is under development.  It performs lazy DAG fusion with nonblocking mode, similar to AppGrB. It executes this DAG, however, using the Suite\-Sparse GraphBLAS rather than generating code directly. Hence, the optimizations that can be applied are limited.

Recent work in tensor compilers~\cite{10.1145/3579990.3580020,10.1145/3133901} has developed techniques for fusing operations via structured iteration, where the compiler understands sparsity structure.  While these techniques have yet to be applied to GraphBLAS operations, we expect that our work can be expanded to leverage some of these techniques.

\section{Conclusions and Future Work}

We used multi-stage programming to implement nonblocking execution in AppGrB, an implementation of  GraphBLAS in Julia.  In this paper, we present our work-in-progress on AppGrB, currently restricted to the subset of the GraphBLAS sufficient to support the LAGraph PageRank algorithm.   The non-blocking AppGrB was faster than the blocking version for all thread counts, validating our general approach to nonblocking execution.  However, it did not provide speedup over the optimized non-blocking SuiteSparse baseline, likely due to inefficiencies in our generated code.  In the future, we plan to extend nonblocking to the rest the GraphBLAS, and we suspect that more complex operations may offer more opportunities for fusion, and thus more speedup for nonblocking execution.  Improving the efficiency of generated code will also help narrow the gap, as demonstrated by our hand-generated C++ code (i.e., AppGrB C++ stub), which is significantly faster than SuiteSparse at high thread counts.

The multi-stage programming approach from Strymonas~\cite{Kiselyov2017}, which influenced our design, describes a limitation in one corner case which should not occur in GraphBLAS.  However, GraphBLAS may pose its own challenges.  For example, select operations might be difficult to optimize using estimated fill ratios.  We suspect,  however, that our method tree representation will support algorithmic optimization for matrix-chain multiplication~\cite{cormen2022} and that we can utilize tile-based fusion~\cite{yzelmanNonblocking2022} and dispatch to optimized libraries~\cite{kimmerer2024} to achieve higher performance.

%\pascal{Does anybody know off the top of their hat what that algorithm for multiplying three matrices is, or which other example we could use here?}

%\bab{A popular contemporary example is FlashAttention (https://arxiv.org/abs/2205.14135), which is an optimized, fused algorithm for computing $QK^{T}V$ where $Q$, $K$, and $V$ are all tall and skinny.  It avoids materializing the intermediate product $QK^{T}$, computing one block of it at a time on demand to multiply by $V$. (It also does some clever math to perform a softmax by row on $QK^{T}$, but that's less relevant here.)  Not sure if you want to mention this, but we could add it to future work.  I'm almost certain you'd need a specific transformation pass on the method tree for this; I don't think the multi-stage programming skeletons would be enough.}

%Future work includes ``happens-before'' relationship, wait(COMPLETE) vs.\ wait(MATERIALIZE); completing the library; combining with other fusion techniques; ...

%%%%%. We don't have room for acknowledgements for HPEC
%\section*{Acknowledgments and Disclaimers}
%
%We thank the members of the GraphBLAS forum.
%

\bibliographystyle{plain}
\bibliography{references}

\scriptsize
\noindent
\newline Optimization Notice: Software and workloads used in
performance tests may have been optimized for performance only on
Intel microprocessors.  Performance tests, such as SYSmark and
MobileMark, are measured using specific computer systems,
components, software, operations and functions.  Any change to any
of those factors may cause the results to vary.  You should
consult other information and performance tests to assist you in
fully evaluating your contemplated purchases, including the
performance of that product when combined with other products.
For more information go to \url{http://www.intel.com/performance}.

\noindent Intel, Xeon, and Intel Xeon Phi are trademarks of Intel Corporation in the U.S. and/or other countries.

\normalsize

\end{document}